\documentclass[twocolumn]{aastex631}

\newcommand{\fesc}{f_\mathrm{esc}^\mathrm{ion}}

\usepackage{url}

\begin{document}

\shorttitle{
A Galaxy with an Extremely Blue UV Continuum
}

\shortauthors{Yanagisawa et al.}

\title{
A Galaxy with an Extremely Blue UV Slope $\beta=-3$ at $z=9.25$ Identified by JWST Spectroscopy:\\
Evidence for a Weak Nebular Continuum and Efficient Ionizing Photon Escape?
}

\author[0009-0006-6763-4245]{Hiroto Yanagisawa}
\affiliation{Institute for Cosmic Ray Research, The University of Tokyo, 5-1-5 Kashiwanoha, Kashiwa, Chiba 277-8582, Japan}
\affiliation{Department of Physics, Graduate School of Science, The University of Tokyo, 7-3-1 Hongo, Bunkyo, Tokyo 113-0033, Japan}

\author[0000-0002-1049-6658]{Masami Ouchi}
\affiliation{National Astronomical Observatory of Japan, National Institutes of Natural Sciences, 2-21-1 Osawa, Mitaka, Tokyo 181-8588, Japan}
\affiliation{Institute for Cosmic Ray Research, The University of Tokyo, 5-1-5 Kashiwanoha, Kashiwa, Chiba 277-8582, Japan}
\affiliation{Department of Astronomical Science, SOKENDAI (The Graduate University for Advanced Studies), 2-21-1 Osawa, Mitaka, Tokyo, 181-8588, Japan}
\affiliation{Kavli Institute for the Physics and Mathematics of the Universe (WPI), University of Tokyo, Kashiwa, Chiba 277-8583, Japan}

\author[0000-0003-2965-5070]{Kimihiko Nakajima}
\affiliation{National Astronomical Observatory of Japan, National Institutes of Natural Sciences, 2-21-1 Osawa, Mitaka, Tokyo 181-8588, Japan}

\author[0000-0002-6047-430X]{Yuichi Harikane}
\affiliation{Institute for Cosmic Ray Research, The University of Tokyo, 5-1-5 Kashiwanoha, Kashiwa, Chiba 277-8582, Japan}

\author[0000-0001-7201-5066]{Seiji Fujimoto}
\affiliation{Department of Astronomy, The University of Texas at Austin, Austin, TX 78712, USA}
\affiliation{Cosmic Dawn Center (DAWN), Denmark}
\affiliation{Niels Bohr Institute, University of Copenhagen, Lyngbyvej 2, DK2100 Copenhagen Ø, Denmark}

\author[0000-0001-9011-7605]{Yoshiaki Ono}
\affiliation{Institute for Cosmic Ray Research, The University of Tokyo, 5-1-5 Kashiwanoha, Kashiwa, Chiba 277-8582, Japan}

\author[0009-0008-0167-5129]{Hiroya Umeda}
\affiliation{Institute for Cosmic Ray Research, The University of Tokyo, 5-1-5 Kashiwanoha, Kashiwa, Chiba 277-8582, Japan}
\affiliation{Department of Physics, Graduate School of Science, The University of Tokyo, 7-3-1 Hongo, Bunkyo, Tokyo 113-0033, Japan}

\author[0009-0000-1999-5472]{Minami Nakane}
\affiliation{Institute for Cosmic Ray Research, The University of Tokyo, 5-1-5 Kashiwanoha, Kashiwa, Chiba 277-8582, Japan}
\affiliation{Department of Physics, Graduate School of Science, The University of Tokyo, 7-3-1 Hongo, Bunkyo, Tokyo 113-0033, Japan}

\author[0000-0002-1319-3433]{Hidenobu Yajima}
\affiliation{Center for Computational Sciences, University of Tsukuba, Ten-nodai, 1-1-1 Tsukuba, Ibaraki 305-8577, Japan}

\author[0000-0002-0547-3208]{Hajime Fukushima}
\affiliation{Center for Computational Sciences, University of Tsukuba, Ten-nodai, 1-1-1 Tsukuba, Ibaraki 305-8577, Japan}

\author[0000-0002-5768-8235]{Yi Xu}
\affiliation{Institute for Cosmic Ray Research, The University of Tokyo, 5-1-5 Kashiwanoha, Kashiwa, Chiba 277-8582, Japan}
\affiliation{Department of Astronomy, Graduate School of Science, The University of Tokyo, 7-3-1 Hongo, Bunkyo, Tokyo 113-0033, Japan}

\begin{abstract}
We investigate UV continuum slopes $\beta$ of 863 galaxies at $4<z<14$ using archival JWST/NIRSpec PRISM spectra obtained from major JWST GTO, ERS, and GO programs, including JADES, CEERS, and UNCOVER. Among these galaxies, we identify a remarkable galaxy at $z=9.25$, dubbed EBG-1, with a significantly blue UV slope $\beta=-2.99\pm0.15$, unlike the rest of the galaxies that exhibit red continua or ambiguous blue continua hindered by large uncertainties. We confirm that the $\beta$ value negligibly changes by the data reduction and fitting wavelength ranges for UV emission/absorption line masking. The extreme blue slope, $\beta=-3.0$, rules out significant contributions from dust extinction or AGN activity. Comparing with stellar and nebular emission models, we find that such a blue UV slope cannot be reproduced solely by stellar models even with very young, metal-poor, or top-heavy contiguous star formation associated with strong nebular continua making the UV slopes red, but with a high ionizing photon escape fraction, $\fesc\gtrsim0.5$, for a weak nebular continuum. While the H$\beta$ emission line is not detected, likely due to the limited sensitivity of the spectrum, we find moderately weak [O\textsc{iii}]$\lambda\lambda$4959,5007 emission lines for the given star-formation rate ($3\, \mathrm{M_\odot}$ yr$^{-1}$) and stellar mass ($10^{8.0} \, \mathrm{M_\odot}
$) that are about three times weaker than the average emission lines, again suggestive of the high ionizing photon escape fraction, $\fesc\sim0.7$ or more. EBG-1 would provide crucial insights into stellar and nebular continuum emission in high-redshift galaxies, serving as an example of the ionizing photon escaping site at the epoch of reionization.
\end{abstract}


\section{Introduction} \label{sec:intro}
The first galaxies initiate star formation in the early universe, with young massive stars exhibiting blue UV continua and producing substantial amounts of ionizing photons.
Although the ionizing photons are mainly used to ionize the interstellar medium, a fraction of the ionizing photons escape from the galaxy and ionize the intergalactic medium, driving the cosmic reionization. The escape of the hydrogen ionizing photons is characterized by an escape fraction, $\fesc$, which is the key quantity to understand how the ionizing photons of the galaxies contribute to the cosmic reionization.

The UV continuum slope $\beta$ ($f(\lambda) \propto \lambda^\beta$) is an important indicator of the production and escape of ionizing photons. Young stellar populations produce $\beta<-2$, while dust extinction and active galactic nucleus (AGN) provide a red ($\beta\gtrsim-2$) UV continuum (e.g., \citealt{Bouwens+12, Finkelstein+12}). Typically the $\beta$ values are larger than $\sim-2.6$ (\citealt{Chisholm+22}), because an intrinsically blue UV continuum intensely ionizes the nebula, leading to a significant contribution from a nebular continuum, which has $\beta \gtrsim -2$ \citep{Katz+24, Cameron+24, Narayanan+2025}. 

However, without the nebular continuum, the $\beta$ values can be as low as $-3.4$ (e.g., \citealt{Bouwens+2010}). This situation is achieved if the escape fraction of the ionizing photon $\fesc$ is large, because in that case the nebula is less ionized and the nebular continuum has less contribution to the UV spectrum \citep{Zackrisson+17}
. It is thus important to search for galaxies with $\beta<-2.6$ because such blue galaxies may have extremely high $\fesc$, which contribute to the cosmic reionization \citep{Dottorini+2024}.

Because galaxies with such high $\fesc$ are probably rare \citep{Leitet+2013, Matthee+2017, Marques-Chaves+2022, Marques-Chaves+2024}, it is necessary to search large sample of galaxies. Photometric studies are conducted with JWST data by \cite{Topping+22, Topping+24}, \cite{Morales+2024}, and \cite{Cullen+2023, Cullen+2024}, although deriving $\beta$ using photometry suffers from contamination by emission lines. The accurate measurement of $\beta$ requires the high quality spectroscopic data. 

In this work, we search a large spectroscopic sample of galaxies
provided by the DAWN JWST Archive for a galaxy with a blue UV slope, and report a galaxy at $z=9.25$ that have an extremely blue UV slope. In Section \ref{sec:sample}, we describe the sample and method of UV slope measurements. In Section \ref{sec:results}, we present the results the extremely blue object. We discuss physical origins of the extremely blue UV continuum in Section \ref{sec:discussion}. Section \ref{sec:summary} summarizes our results.

\begin{figure}
    \centering
    \includegraphics[width=1\linewidth]{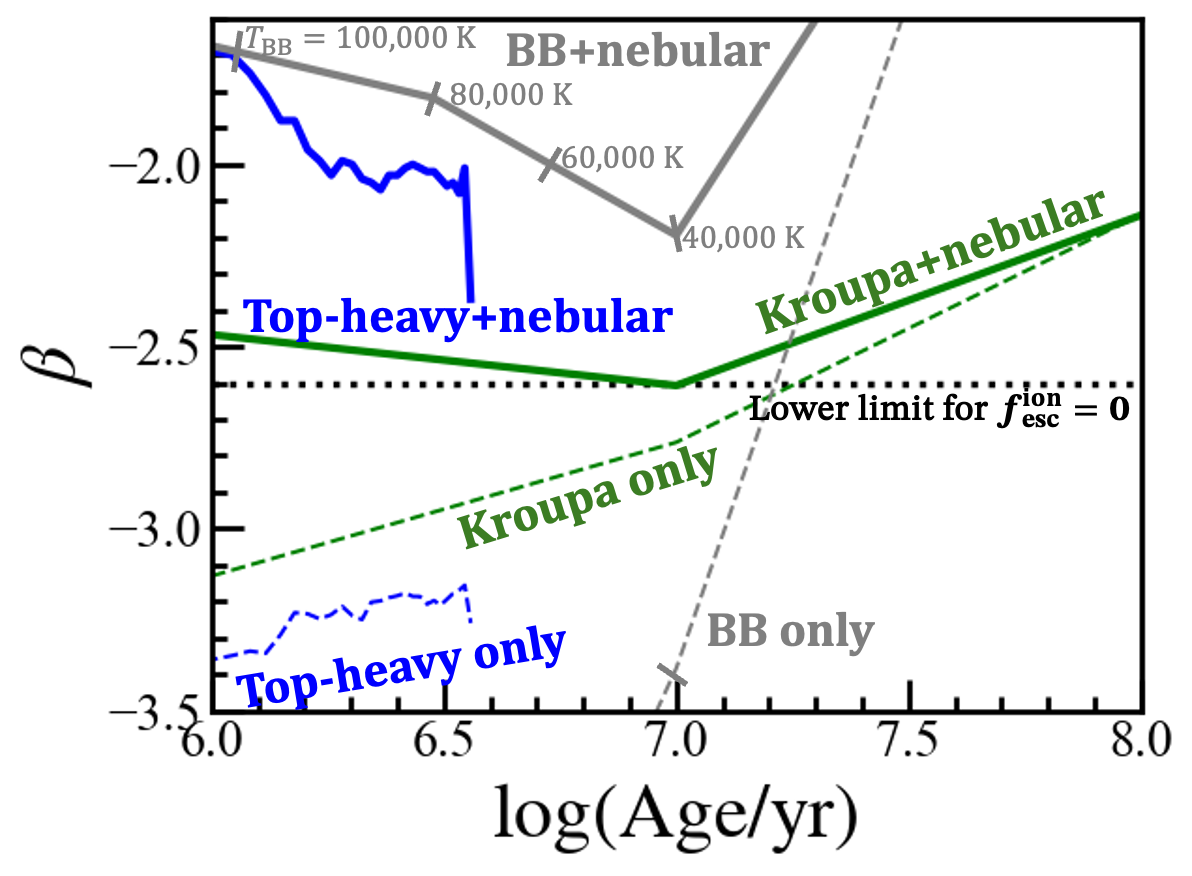}
    \caption{UV slope $\beta$ as a function of age calculated with Cloudy. The blue, green, and gray lines present the incident radiations of top-heavy IMF \citep{Zackrisson+11}, Kroupa IMF \citep{Kroupa+01}, and blackbody, respectively. The solid lines denote the models with $\fesc = 0$ (i.e., the ionizing photons are completely consumed to ionize the nebula), while the dashed lines represent the models with $\fesc = 1$ (i.e., the ionizing photons are not used to ionize the nebula). For the blackbody models, the blackbody temperatures are converted into the age using the typical lifetimes of stars having the same temperature.}
    \label{fig:beta-age}
\end{figure}

\section{Search for Blue UV Slope Objects}\label{sec:sample}
\subsection{Lower Limit of $\beta$ for $\fesc=0$}\label{sec:lowerlimit}
We first define a quantitative criterion for a blue UV continuum. We calculate $\beta$ assuming $\fesc=0$ using Cloudy version 23.01 \citep{Cloudy23} with incident radiations from Kroupa IMF \citep{Kroupa+01} with a mass range of $0.1-100 \, \mathrm{M_\odot}$ provided by BPASS v2.2.1 \citep{bpass221}, top-heavy IMF with a mass range of $50-500 \, \mathrm{M_\odot}$ taken from the Yggdrasil Pop III.1 model \citep{Zackrisson+11}, and blackbody. 
We assume a hydrogen density of $n_\mathrm{e} = 10^{2} \, \mathrm{cm^{-3}}$, nebular metallicity of $\log Z_\mathrm{neb}/\mathrm{Z_\odot} = -2$, number of ionizing photons $Q(\mathrm{H})=10^{50} \, \mathrm{s^{-1}}$, and inner radius of $10^{14} \, \mathrm{cm}$. We assume stellar metallicity of $\log Z_\mathrm{star}/\mathrm{Z_\odot} = -2$ for the Kroupa IMF model, while $Z_\mathrm{star}=0$ is used for top-heavy IMF. Figure \ref{fig:beta-age} shows the evolution of $\beta$ values as a function of stellar age. Although the incident radiation of the top-heavy IMF is bluer than that of the Kroupa IMF, the top-heavy IMF+nebular continuum model is redder than the Kroupa+nebular continuum model. This is because the intrinsically blue incident radiation in top-heavy IMF model intensely ionize the nebula, which leads to a significant contribution from the red nebular continuum. One can also see that the lower limit of $\beta$ value for $\fesc=0$ is $-2.6$, while the $\beta$ value reaches as low as $-3.4$ if nebular continuum is not included. We thus define a criterion for a blue UV continuum as $\beta=-2.6$.

\subsection{DJA Data}\label{sec:dja}
We use PRISM/CLEAR spectra provided by the DAWN JWST Archive (DJA) \footnote{\url{https://dawn-cph.github.io/dja/index.html}}, which compile the major JWST GTO, ERS, and GO programs 
including ERS 1345 (CEERS; PI: Finkelstein), DDT 2756 (PI: Chen), DDT 2750 (CEERS; PI: Arrabal Haro), GTO 1180 (PI: Eisenstein), GTO 1181 (PI: Eisenstein), GTO 1210 (PI: Luetzgendorf), GTO 1211 (PI: Isaak), GTO 1286 (PI: Luetzgendorf), DDT 6541 (PI: Egami), GO 1433 (PI: Coe),  GO 1747 (PI: Roberts-Borsani), GO 1810 (PI: Belli), GO 1871 (PI: Chisholm), GO 2110 (PI: Kriek), GO 2198 (PI: Barrufet), GO 2561 (UNCOVER; PI: Labbe), GO 2565 (PI: Glazebrook), DDT 2767 (PI: Kelly), ERO 2736 (PI: Pontoppidan),  GO 3215 (PI: Eisenstein), GO 4233 (PI: de Graaff), GO 4246 (PI: Abdurro'uf), DDT 4446 (PI: Frye), and DDT 4557 (PI: Yan). 
We select the 2056 objects within the redshift range of $4.1<z<13.4$, whose UV continua are securely covered by NIRSpec. The spectra in the DJA are reduced with {\tt\string msaexp} (\citealt{msaexp}; \citealt{Heintz2024, deGraaff+2024}). 

\begin{figure}
    \centering
    \includegraphics[width=1\linewidth]{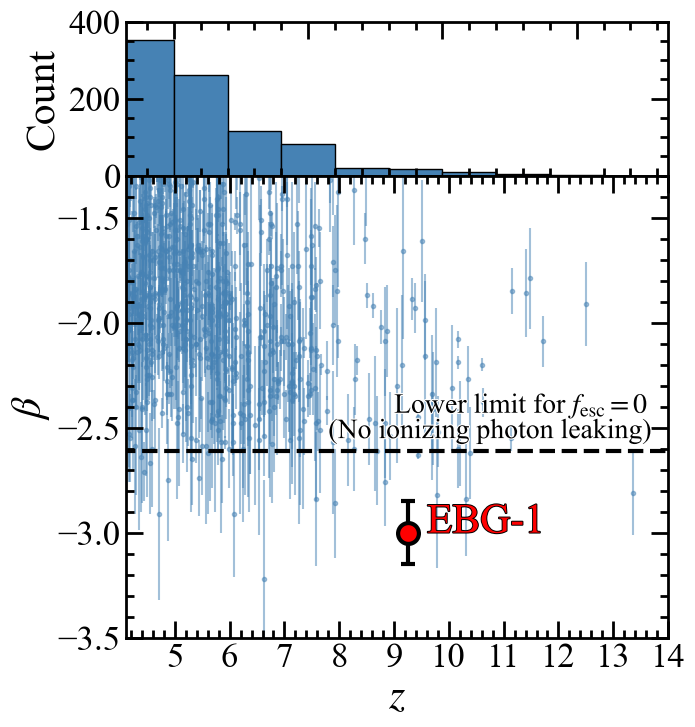}
    \caption{(Top) Redshift distribution of our sample. (Bottom) Fitted $\beta$ value as a function of redshift. The blue points represent the galaxies in our sample. The $\beta$ values with large error ($>0.5$) are omitted in this figure. The red point indicates EBG-1. The dotted line denote the lower limit of $\beta$ for $\fesc=0$.}
    \label{fig:beta-z}
\end{figure}

\begin{figure*}
    \centering
    \includegraphics[width=1\linewidth]{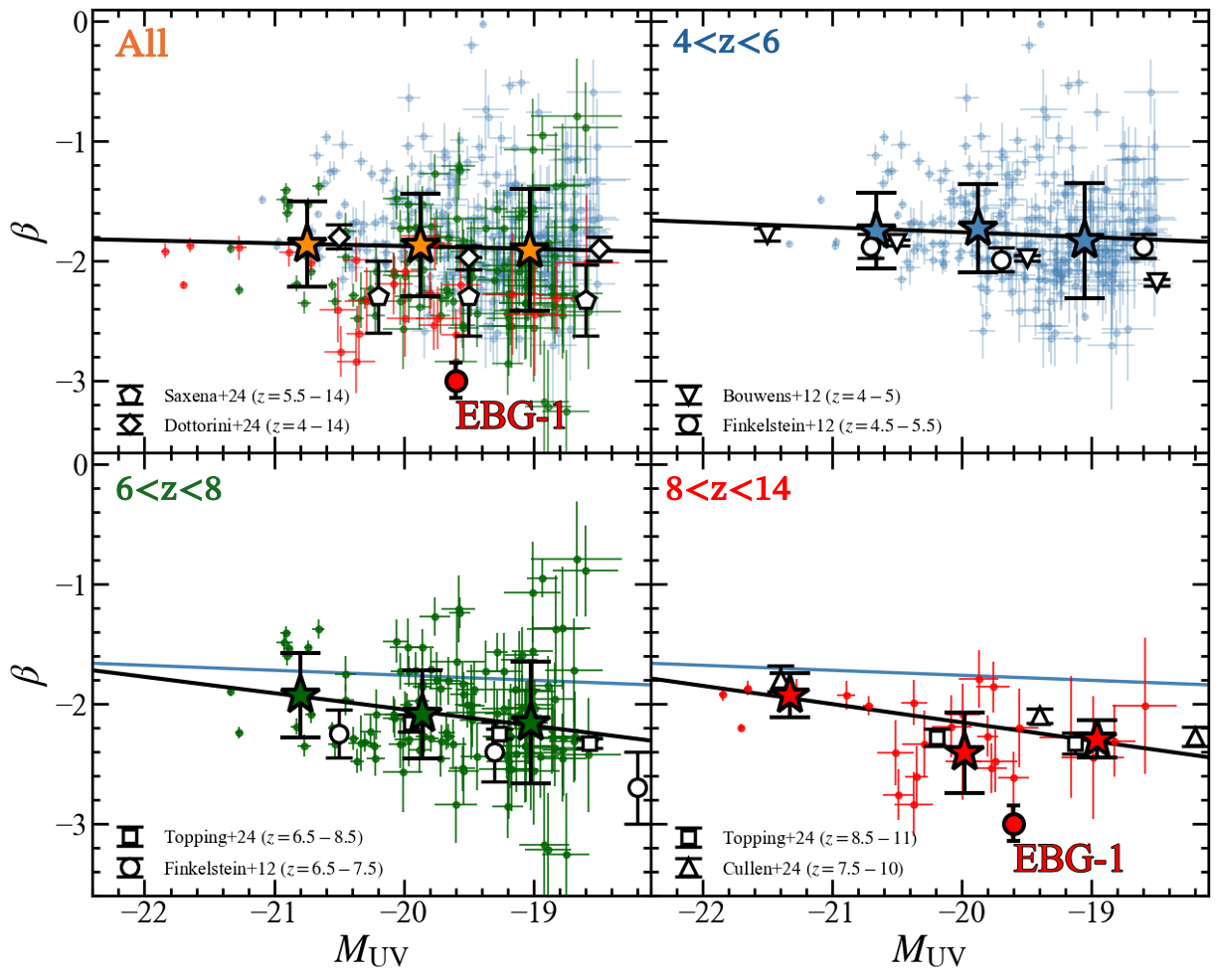}
    \caption{Relation between $\beta$ and $M_\mathrm{UV}$. From top-left to bottom-right, each panel shows galaxies at $z=4-14$, $4-6$, $6-8$, and $8-14$. The blue, green, and red points represent galaxies at $4-6$, $6-8$, and $8-14$, respectively, while the large red points represent EBG-1. The stars and error bars denote the median and standard deviation in $M_\mathrm{UV}$ bins. The black lines show the linear fit to the median points in each redshift bin, while the blue lines in the bottom two panels denote the linear fit at $z=4-6$. The open symbols represent the previous studies \citep{Bouwens+12, Finkelstein+12, Saxena+2024, Dottorini+2024, Cullen+2024, Topping+24}.}
    \label{fig:beta-Muv}
\end{figure*}

\subsection{UV Slope Measurements}\label{sec:macs}
To derive $\beta$, we fit a spectra with a function

\begin{equation}\label{eq:beta}
    f(\lambda) = A \lambda^{\beta},
\end{equation}
where $A$ is a constant for normalization and $\lambda$ is wavelengths. We employ a Markov Chain Monte Carlo (MCMC) method for the fitting using emcee \citep{Foreman-Mackey+2013_emcee}. We minimize a likelihood function

\begin{equation}
    \log (\mathcal{L}) = - \frac{1}{2} \sum_{\lambda} \left[\left(\frac{f_{\mathrm{mod}}(\lambda) - f_{\mathrm{obs}}(\lambda)}{\sigma(\lambda)}\right)^2 + \log(\sigma(\lambda)^2)\right],
\end{equation}
where $f_\mathrm{mod}$ and $f_\mathrm{obs}$ are the model and observed flux, respectively, and $\sigma(\lambda)$ is the 1$\sigma$ error of the flux. The best-fit parameters and their 1$\sigma$ uncertainties are determined by taking the medians and 68 percentiles of the posterior distributions, respectively. We use a rest-frame wavelength range of $1268-2580\,$ \AA \, for the fitting, following the fitting range presented by \cite{Calzetti+94}. The measured $\beta$ values are shown in the bottom panel of Figure \ref{fig:beta-z}. Here we exclude galaxies with $\beta$ errors larger than 0.5, because such galaxies have large uncertainties in their spectra possibly due to the low sensitivity. This results in the sample of 863 galaxies, which is referred to as our sample. Our error estimates are comparable to those of \cite{Dottorini+2024}, but larger than those of \cite{Saxena+2024}. This is because \cite{Saxena+2024} exclude galaxies whose $\beta$ uncertainties are larger than 5\%, which is stricter than our criterion. We also present $\beta$ as a function of UV magnitude $M_\mathrm{UV}$ in Figure \ref{fig:beta-Muv}. Here we exclude galaxies with $M_\mathrm{UV}>-18.5$, which could be influenced by the low completeness. We only plot the galaxies whose magnification factors are available or negligible. Our result is broadly consistent with previous studies, which show weak anti-correlation between $\beta$ and $M_\mathrm{UV}$ \citep{Saxena+2024, Dottorini+2024, Cullen+2024}. One can also see the trend of lower $\beta$ at high redshift, again consistent with the previous work.

\begin{figure*}
    \centering
    \includegraphics[width=1\linewidth]{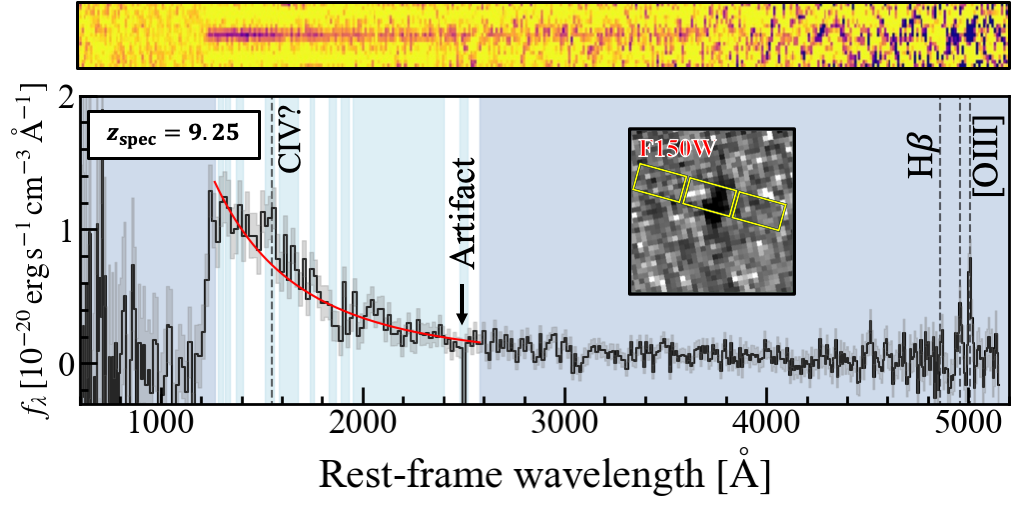}
    \caption{Top and bottom panels show the 2D and 1D NIRSpec spectra of EBG-1 reduced by DJA, respectively. In the bottom panel, the black histogram and gray shaded regions represent the spectrum and its 1$\sigma$ uncertainty, respectively. The red line presents the best-fit UV slope derived with the \cite{Calzetti+94} fitting windows. The dark blue shaded regions are the regions that are not used for the UV slope fitting. The light blue shaded regions denote the masks presented by \cite{Calzetti+94} and the mask for the artifact at $2500\,$\AA, which are not used for fitting. The dotted lines present the positions of emission lines. The F150W image and slit position are presented in the inset.}
    \label{fig:spec}
\end{figure*}

Most of the galaxies show $\beta>-2.6$, which is larger than the lower limit for $\fesc=0$ (Figure \ref{fig:beta-age}). Among our sample, we find one galaxy at $z=9.25$, showing $\beta=-2.99\pm0.15$, which is smaller than $-2.6$ beyond the 2$\sigma$ level. Hereafter, we refer to this object as extremely blue galaxy 1 (EBG-1). EBG-1 is originally identified as MACS0647-z9-20158 at $z_\mathrm{phot}=9.5$ by the program GO 1433 (PI: Coe; \citealt{McLeod+24}), and then spectroscopically observed by the same program GO 1433 as a filler target for MACS0647-JD. The NIRCam images and spectra of EBG-1 are shown in Figure \ref{fig:spec}.  

\begin{figure}
    \centering
    \includegraphics[width=1\linewidth]{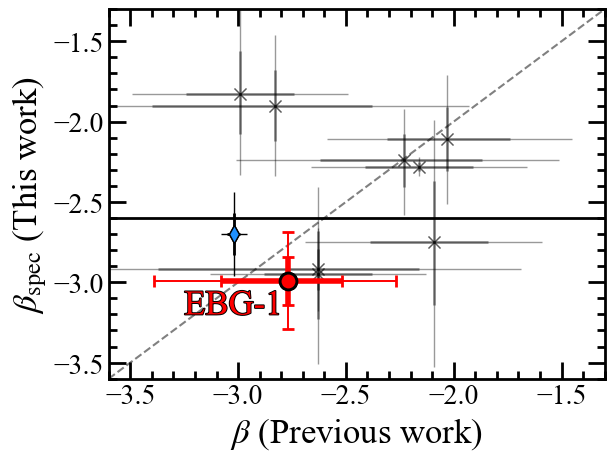}
    \caption{Comparison of $\beta$ measurements in this work and previous work. The red circle represents the values for EBG-1, whose $\beta$ values in the previous works are taken from the photometric measurement of \cite{Cullen+2024}. The crosses denote the $\beta$ values taken from the photometric measurements of \cite{Cullen+2024}, while the blue diamond represents those taken from the spectroscopic value of \cite{Saxena+2024}. The thick and thin error bars represent 1 and 2$\sigma$, respectively. The black solid line present the lower limit of $\beta$ for no ionizing photon escape. The gray dashed line denotes the line of equality. }
    \label{fig:betacomparison}
\end{figure}

Recently, \cite{Saxena+2024} have found six galaxies at $5.5<z<8$ showing $\beta\sim-3$ from the spectroscopic sample obtained from the JADES survey. One out of six galaxies, JADES-GS-210003 at $z=5.779$, is also included in our sample (as of version 2 of the DJA spectra, the other five galaxies were not publicly available). Using the same fitting range as \cite{Saxena+2024}, we obtain $\beta=-2.70\pm0.13$ for JADES-GS-210003 (Figure \ref{fig:betacomparison}). There is a $\sim2\sigma$ difference between our and \cite{Saxena+2024} of the $\beta$ values. This is probably because \cite{Saxena+2024} conduct a sigma-clipping method to exclude outlying pixels. JADES-GS-210003 is not selected in our study because we cannot distinguish from $\beta=-2.6$ at the $2\sigma$ level. On the other hand, EBG-1 is not selected by \cite{Saxena+2024}, as EBG-1 falls in the MACS0647 lensing field that is not covered in the sample of \citet{Saxena+2024}.

We also compare our $\beta$ values with those in previous studies in Figure \ref{fig:betacomparison}. We plot photometric $\beta$ measurements of \cite{Cullen+2024}, who conduct $\beta$ measurements for the sample galaxies taken from NGDEEP, JADES DR1, UNCOVER, and those from \cite{McLeod+24}. \cite{Cullen+2024} also derive $\beta$ for EBG-1, which show good agreement with our measurement.

\subsection{Observations and Data of EBG-1}
In the previous section, we identify EBG-1  from the DJA spectra. In Section \ref{sec:obs}, we first describe how EBG-1 was observed. We next conduct the photometry for EBG-1 to confirm whether the photometry is consistent with spectrum in Section \ref{sec:photo}. We then independently performed data reduction to verify whether the $\beta$ value changes with different data reduction procedures, as described in Section \ref{sec:reduction}.

\subsubsection{Observations}\label{sec:obs}
MACS0647 lensing field was observed with JWST/NIRCam in January 8th 2023 in GO 1433 (PI: Coe) targeting MACS0647-JD \citep{Coe+2013, Hsiao+2023}. EBG-1 is falling on the footprints of this NIRCam observations, which is photometrically identified at R.A.$=$06:47:36.95 and Decl.$=$+70:14:34.69 by \cite{McLeod+24}. 

EBG-1 was then spectroscopically observed with JWST/NIRSpec as a filler target for MACS0647-JD in GO 1433 in February 20th, 2023. The observations were performed with PRISM/CLEAR ($R\sim100$) for a total exposure time of 2200 s. The data were reduced with {\tt\string msaexp} by DJA. For details of the reduction, see \cite{Heintz2024}. The spectrum reduced by DJA is shown in Figure \ref{fig:spec}.

\subsubsection{Photometry}\label{sec:photo}
Calibrated NIRCam data are collected from DJA. We conduct aperture photometry with $0.^{\prime\prime}35$ aperture size, which are shown in Table \ref{tab:properties}. The errors are estimated by conducting photometry within a $0.^{\prime\prime}80$ annulus and taking the standard deviations. An apparent magnitude of F200W is derived as 27.7, which is consistent with that derived by \cite{McLeod+24}.

\subsubsection{Reanalysis of EBG-1 spectra}\label{sec:reduction}
\begin{figure*}
    \centering
    \includegraphics[width=1\linewidth]{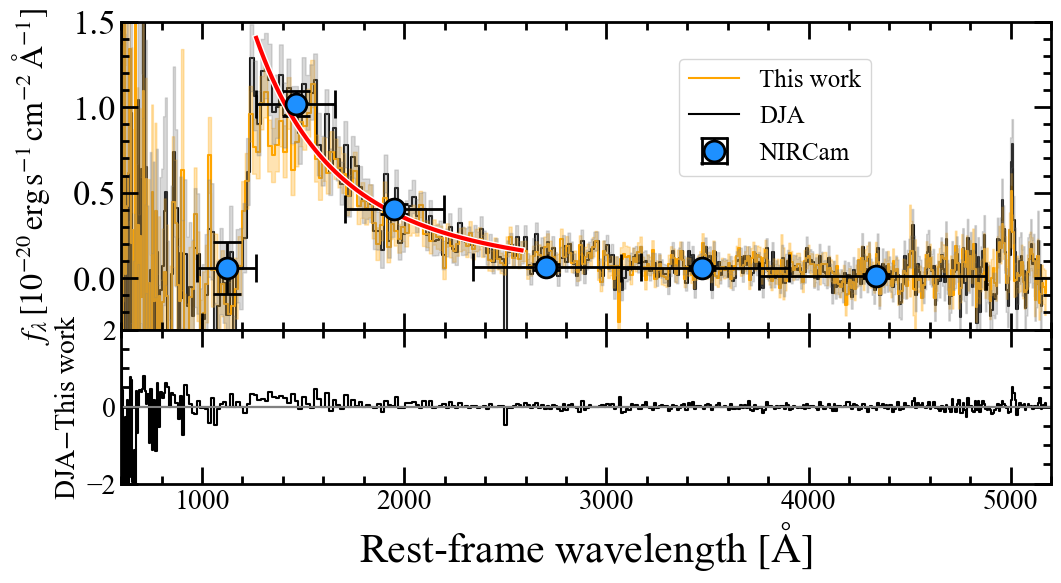}
    \caption{Comparison of NIRSpec spectrum of EBG-1 reduced in this work and DJA. (Top) The yellow histogram and shaded region represent the spectrum reduced in this work and its 1$\sigma$ error, respectively. The blue points denote the NIRCam photometry. The other symbols are the same as Figure \ref{fig:spec}. (Bottom) Difference between the spectra reduced by DJA and in this work.}
    \label{fig:spec2}
\end{figure*}

We performed the data reduction in this work following the method described in \cite{Nakajima+2023}. Starting from the level 1 products provided by MAST, we executed Spec2 and Spec3 pipelines using Python library {\tt\string jwst} (ver. 1.16.1; \citealt{bushouse_2024_14153298_jwst_pipeline}). The reference files stored in the latest pmap file of {\tt\string jwst\_1299.pmap} were used. The pathloss corrections were conducted by comparing the position of the source and MSA shutter, where we assumed that the source was point-like. We then combined 2D spectra by median-stacking to reduce the effect of hot pixels, with extractions of 3 pixels in the spatial direction. For more details, see \cite{Nakajima+2023}. In Figure \ref{fig:spec2} we present the spectrum reduced in this work, which is consistent with both of the spectrum reduced by DJA and the NIRCam photometry.

The spectrum of EBG-1 shows [O \textsc{iii}] $\lambda\lambda$4959, 5007 and tentative C \textsc{iv} $\lambda\lambda$1548, 1550 emission lines. We measure flux values and $3\sigma$ upper limits of emission lines by integrating the flux and error in each wavelength bin (Table \ref{tab:flux}). The C \textsc{iv} emission might be associated with an AGN, although the extremely blue UV slope disfavors the contribution from an AGN, whose dust content reddens the spectrum \citep{Francis+1991}.

\section{Results}\label{sec:results}
\subsection{Confirmation of the Blue UV Slope}
To check whether the extremely blue UV continuum of EBG-1 is caused by systematics or not, we adopt three fitting methods: 1) conventional \cite{Calzetti+94} windows, 2) \cite{Saxena+2024} windows, which are slight modifications of \cite{Calzetti+94} ones, 3) using the whole wavelength range without masking, 4) simply avoiding the possible C \textsc{iv} emission line, and 5) avoiding the possible Ly$\alpha$ damping wing at $<1340$ \AA\, as explored by \cite{Dottorini+2024} (note that we mask out at $2480-2520\,$\AA \, contaminated by the artifact). For the spectrum reduced by DJA, each fitting method gives $\beta = -3.03\pm0.22, -3.17\pm0.23, -2.99\pm0.15, -3.08\pm0.15$, and $-3.33\pm0.16$, respectively, all of which are smaller than $\beta=-2.6$ at the $\gtrsim2\sigma$ level. 

We derive $\beta$ also for the spectrum reduced in this work in the same manner as described above. The fitting methods of 1), 2), 3), 4) and 5) yield $\beta=-2.96\pm0.19$, $\beta=-3.07\pm0.21$, $-2.92\pm0.13, -2.79\pm0.14$, and $-3.16\pm0.15$, respectively. Although the significance levels are slightly lower than the $\beta$ values from the DJA spectrum, the results of the extremely blue UV slope of EBG-1 does not significantly change. We thus conclude that the extremely blue UV slope of EBG-1 is not attributed to observational systematics.

\begin{deluxetable}{cc}
    \centering
    \tablecaption{Properties of EBG-1}
    \tablehead{
    \colhead{Quantity} & \colhead{Value}
    }
    \startdata
    R.A. & 06:47:36.95 \\
    Decl. & +70:14:34.69\\
    $z_\mathrm{spec}$ & 9.25 \\
    $\mu$ & $2.75$ \\
    F115W & $1.54\pm3.87 \, \mathrm{nJy}$  \\
    F150W & $44.0\pm3.2 \, \mathrm{nJy}$  \\
    F200W & $31.0\pm2.2 \, \mathrm{nJy}$  \\
    F277W & $9.34\pm2.60 \, \mathrm{nJy}$  \\
    F356W & $14.8\pm2.7 \, \mathrm{nJy}$  \\
    F444W & $4.79\pm3.79 \, \mathrm{nJy}$  \\
    $M_\mathrm{UV}$ & $-19.6^a$ \\
    $\mathrm{SFR}_{\mathrm{UV}}$ &  $3.0 \pm 0.2 \, \mathrm{M_\odot/yr}$ \\
    $\mathrm{SFR_{Prospector}}$ & $5.5 \pm 0.5 \, \mathrm{M_\odot/yr}$ \\
    $\log M_*/\mathrm{M_\odot}$ &  $7.98^{+0.09}_{-0.04}$  \\
    $\tau_\mathrm{dust}$(5500\,\AA) & $0.01^{+0.02}_{-0.01}$ \\
    $\log Z/\mathrm{Z_\odot}$ & $-3.94^{+0.10}_{-0.04}$ \\
    $\log U$ & $-3.25^{+3.03}_{-1.88}$ \\
    \enddata
    \tablecomments{$^a$ The value is corrected for magnification factor.}
    \label{tab:properties}
\end{deluxetable}

\begin{deluxetable}{cc}
    \centering
    \tablecaption{Line fluxes and $3\sigma$ upper limits}
    \tablehead{
    \colhead{Line} & \colhead{Flux ($10^{-18} \mathrm{erg \, s^{-1} \, cm^{-2}}$)}
    }
    \startdata
    Ly$\alpha$                                 & $<1.5$ \\
    C \textsc{iv} $\lambda\lambda$1548, 1550   & $<3.43$ \\ \relax
    [O \textsc{ii}] $\lambda\lambda$3727, 3729 & $<0.57$ \\
    H$\beta$                                   & $<0.99$ \\ \relax
    [O \textsc{iii}] $\lambda$5007             & $1.46 \pm 0.32$ \\
    \enddata
    \tablecomments{}
    \label{tab:flux}
\end{deluxetable}

\subsection{SED Fitting}\label{sec:sedfitting}
\begin{figure}
    \centering
    \includegraphics[width=1\linewidth]{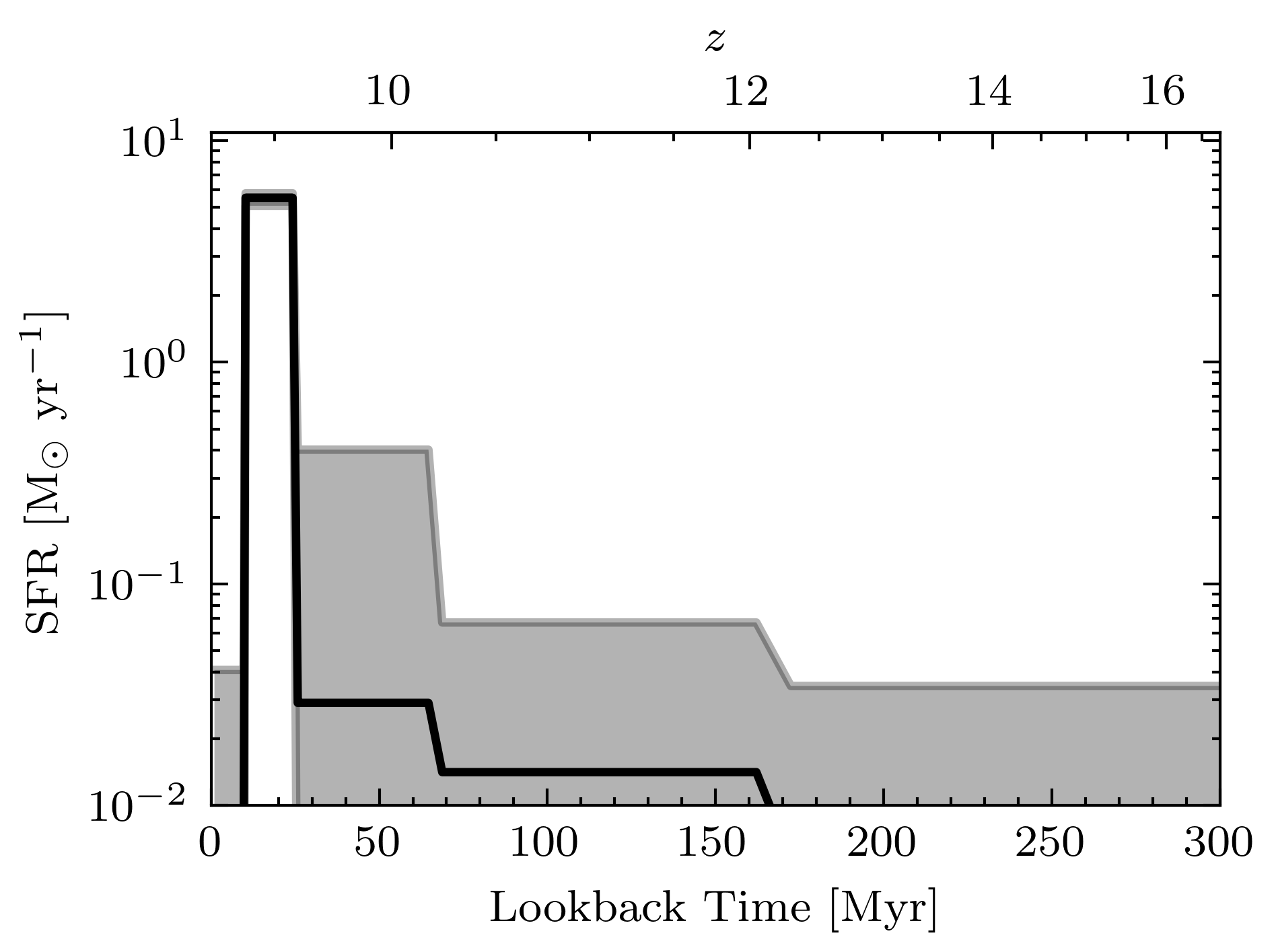}
    \caption{Best-fit SFH for EBG-1. The black line and gray shaded region represent the best-fit and $1\sigma$ error of SFH, respectively.}
    \label{fig:sfh}
\end{figure}

To measure stellar properties, we conduct SED fitting to the NIRCam data presented in Table \ref{tab:properties} using Prospector \citep{prospector}. We use the photometry values that are corrected for magnification factor derived by \cite{McLeod+24}, who utilized the Zitrin-NFW and Zitrin-LTM-Gauss lensing models \citep{Zitrin+2015}. The Binary Population and Spectral Synthesis (BPASS; \citealt{Eldridge+2017_bpass}) model is used for an isochrone library. We apply \cite{Calzetti+2000} dust extinction law. We assume a non-parametric star-formation history (SFH) with five bins that are evenly spaced in logarithmic times between 0 Myr and a look-back time corresponding to $z=30$. The redshift of the source is fixed at $z_\mathrm{spec}=9.25$. We vary the stellar mass $M_*$, metallicity $Z$, optical depth of dust attenuation at 5500 \AA \, $\tau_\mathrm{dust}$(5500\,\AA), and ionization parameter $U$. The best-fit parameters and SFH are shown in Table \ref{tab:properties} and Figure \ref{fig:sfh}, respectively. The results imply the dust-free condition and the recent starburst, which agree with the extremely blue UV continuum of EBG-1. Because $\fesc=0$ is assumed in our SED fitting, the weak emission line feature of EBG-1 indicates a low metallicity, which compensates for the effect of nonzero $\fesc$.

\subsection{Weak Nebular Emission Lines}
\begin{figure}
    \centering
    \includegraphics[width=1\linewidth]{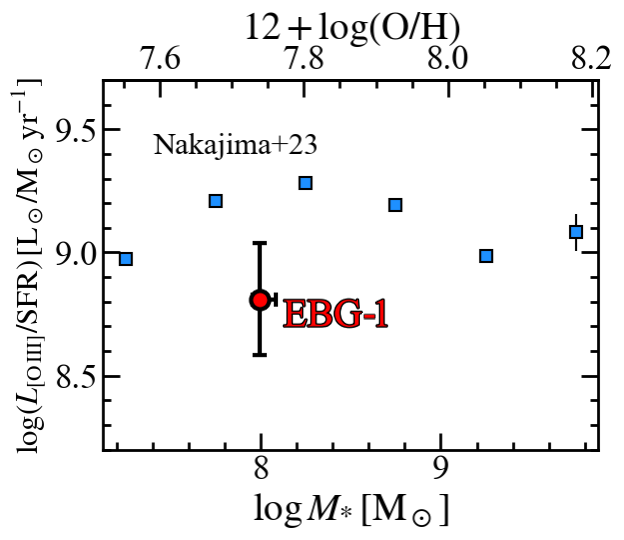}
    \caption{$L_\mathrm{[OIII]}$/SFR as a function of stellar mass. The red circle represents EBG-1. The blue squares denote the average values of galaxies at $4<z<9$ taken from \cite{Nakajima+2023}. The ticks at the top of the figure denote metallicity, which is converted from stellar mass using the mass-metallicity relation at $z=4-10$ derived by \cite{Nakajima+2023}.}
    \label{fig:O3SFR-Mstar}
\end{figure}

The extremely blue UV continuum in EBG-1 may be explained by high $f_\mathrm{esc}$. However, the detection of [O \textsc{iii}] $\lambda\lambda$4959,5007 emission indicates a certain amount of the ionizing photon is used to ionize the nebulae. To quantify the escape of ionizing photons in EBG-1, we plot ratios of luminosity of [O \textsc{iii}] $\lambda$5007 line, $L_\mathrm{[OIII]}$, to star-formation rate (SFR) as a function of stellar mass in Figure \ref{fig:O3SFR-Mstar}. For comparison, we also plot average values calculated from 126 galaxies at $4<z<9$ compiled by \cite{Nakajima+2023}. The SFRs are estimated from UV luminosities of galaxies by using Equation (1) of \cite{Kennicutt+1998_review}. The $L_\mathrm{[OIII]}$/SFR value of EBG-1 is $\sim0.5$ dex smaller than the average value at the same stellar mass, suggesting that [O \textsc{iii}] emission in EBG-1 is $\sim3$ times weaker than the average. This can be interpreted as a result of the escape of ionizing photon without ionizing the nebulae. If we assume $\fesc=0$ for the galaxies in \cite{Nakajima+2023}, the [O \textsc{iii}] emission three times weaker than the average suggests $\fesc\sim0.7$ for EBG-1. However, the $\fesc$ values of the galaxies in \cite{Nakajima+2023} can be larger than zero, in which case our estimate of $\fesc$ becomes larger. Furthermore, if we assume density-bounded nebulae, weak [O \textsc{iii}] emission means an excess of $\mathrm{O^{++}}$ ionizing photons (ionization energy of $35 \, \mathrm{eV}$) compared to the amount of the gas, with $\mathrm{H^+}$ ionizing photon (ionization energy of $13.6 \, \mathrm{eV}$) being even more abundant. In such a situation, the escape fraction of $\mathrm{H^+}$ ionizing photon can be larger than that of $\mathrm{O^{++}}$ ionizing photon.

\begin{figure*}
    \centering
    \includegraphics[width=0.8\linewidth]{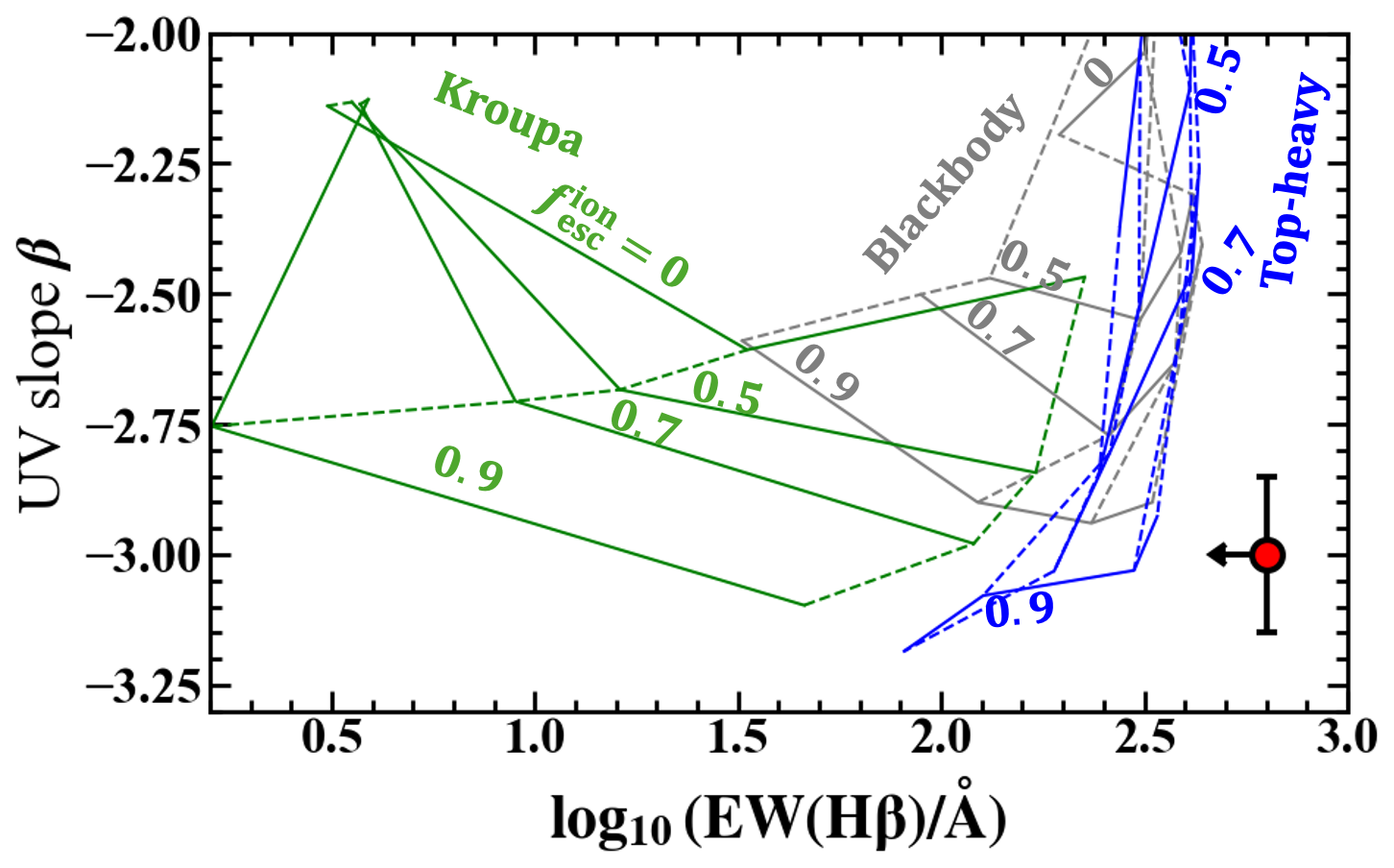}
    \caption{Relation of $\beta$ and EW(H$\beta$). The blue, green, and gray grids denote the same models in Figure \ref{fig:beta-age}. The numbers shown beside the grids represent $\fesc$. The red circle presents the current constraint of $\beta=-2.99\pm0.15$.}
    \label{fig:beta-ew}
\end{figure*}

However, the estimate of $\fesc$ from the [O \textsc{iii}] emission is susceptible to metallicity. To estimate $\fesc$ independently of the assumption of metallicity, we use the $\beta$ and EW(H$\beta$) values of EBG-1. We derive the upper limit of the H$\beta$ flux by summing up the error spectrum in quadrature. We then divide the upper limit of H$\beta$ by the continuum flux density, obtaining $\log\mathrm{(EW(H\beta)/}$\AA)$<2.8$. In Figure \ref{fig:beta-ew}, we compare the $\beta$ and EW(H$\beta$) values with the Cloudy models, which are the same as those in Figure \ref{fig:beta-age} except for varying $\fesc$. In all of the models, the very small $\beta$ values of EBG-1 imply $f_\mathrm{esc}\gtrsim0.5$ (although models with extremely metal-poor or metal-free stellar populations reproduce $\beta\sim-3$ even with $\fesc=0$ (Figure 4 of \citealt{Bouwens+2010}), the detection of [O \textsc{iii}] emission in EBG-1 disfavors this scenario). However, due to the weak upper limit of EW(H$\beta$), it is difficult to constrain the ionizing radiation and $f_\mathrm{esc}$ with current data.

\section{Discussion}\label{sec:discussion}
\begin{figure}
    \centering
    \includegraphics[width=1\linewidth]{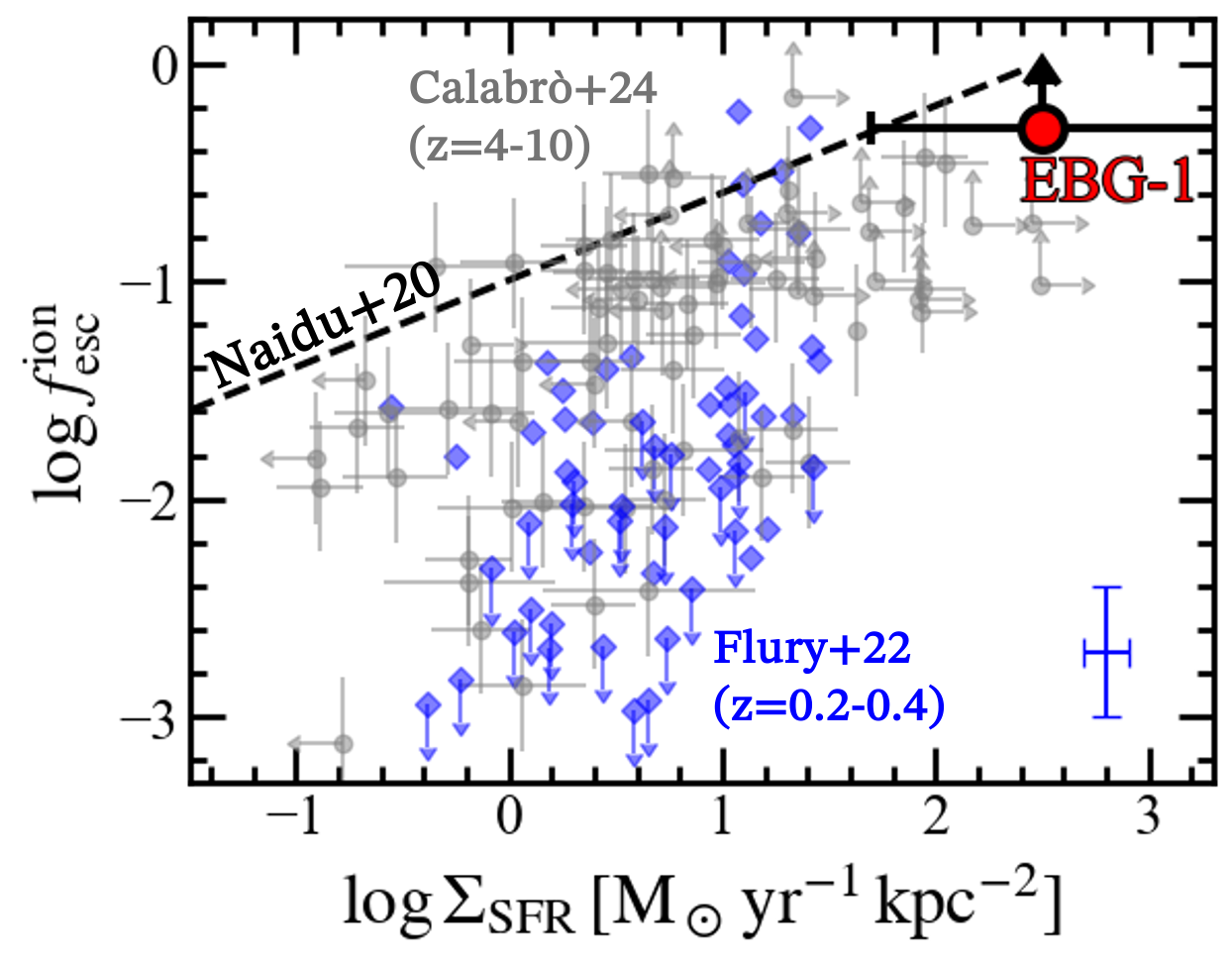}
    \caption{Relation between $\fesc$ and $\Sigma_\mathrm{SFR}$. The red point shows EBG-1. The blue diamonds represent the galaxies obtained from the Low-z Lyman Continuum Survey (LzLCS; \citealt{Flury+2022}), with the blue error bar shown at the bottom right corner representing the typical error. The gray points denote the JWST galaxies at $z=4-10$ presented by \cite{Calabro+2024}. The black dashed line shows the relation $\fesc \propto \left(\Sigma_\mathrm{SFR}\right)^{0.4}$ suggested by \cite{Naidu+2020}.}
    \label{fig:sigmaSFR}
\end{figure}

Although the origin of the high escape fraction remains uncertain, the morphology of EBG-1 (Figure \ref{fig:spec}) may provide a clue. 
We conduct GALFIT \citep{Peng+2002_galfit, Peng+2010_galfit} fitting to the NIRCam F200W image to estimate an effective radius $r_\mathrm{e}$ following the method presented by \cite{Ono+2024}. We obtain $r_\mathrm{e} = 0.04 \, \mathrm{kpc}$, which results in an extremely high SFR surface densities of $\log (\Sigma_\mathrm{SFR}/\mathrm{M_\odot \, yr^{-1} \, kpc^{-2})} = 2.5$. In previous studies, some Lyman continuum leakers show the compact morphologies and high SFR surface densities, pointing out the correlation between $\fesc$ and $\Sigma_\mathrm{SFR}$ \citep{Naidu+2020}. We compare $\fesc$ and $\Sigma_\mathrm{SFR}$ of EBG-1 with previous measurements in Figure \ref{fig:sigmaSFR}. EBG-1 is located on the sequence of the high-$z$ galaxies presented by \cite{Calabro+2024}, while the $\fesc$ and $\Sigma_\mathrm{SFR}$ values of EBG-1 are among the highest in the galaxies presented in previous studies.
Several hydrodynamical simulations suggest that the star formation and supernovae occurring in the compact region produces channels in the ISM through which the ionizing photons can escape (e.g., \citealt{Katz+2018}), which may be the case for EBG-1.

Another possibility may be suggested by a small tail extending towards north-west direction from the main component. The slit covers around the tail of the EBG-1. The tail may be stripped from EBG-1 by galaxy interactions, where the gas could be lacking. We may be looking at this gas-deficient region, where the nebulae are density-bounded or with many holes. In such a case, the center of EBG-1 can be redder than the tail (see also \cite{Schombert+1990}, who claim that tidal tails tend to be bluer than primary galaxies from observations). To examine this scenario, it is necessary to conduct spectroscopy by placing a slit along the north-west direction that covers from the center to the tail of the EBG-1 with spatial extent.

The fraction of the extremely blue galaxy in this work is low compared to the other studies such as \cite{Saxena+2024} and \cite{Dottorini+2024}. This is probably because of the low signal-to-noise ratio of most of the galaxies. Although some galaxies show the best-fit $\beta$ values smaller than $-2.6$ in Figure \ref{fig:beta-z}, the large error bars hinder us from confirming that these galaxies truly have blue UV continua. There may be more galaxies like EBG-1, while deeper spectroscopic observations are required to confirm. However, if galaxies with extremely high $\fesc$ are too abundant, cosmic reionization could complete too early \citep{Munoz+2024}.

The non-detection of H$\beta$ line and the large uncertainty in [O \textsc{iii}] emission lines make it difficult to constrain the $\fesc$ value and type of ionizing source of EBG-1. It is thus necessary to conduct deeper spectroscopy on EBG-1 to definitively confirm the high $\fesc$ and the ionizing source, by which one can utilize EBG-1 as an example to study the process of the significant ionizing photon escape during the epoch of reionization.

\section{Summary}\label{sec:summary}
In this work, we search the large JWST/NIRSpec spectroscopic sample for a galaxy with an extremely blue UV continuum. Among our sample consisting of the 863 galaxies at $4 < z < 13$ taken from the major JWST GTO, ERS, and GO programs, we identify EBG-1, a galaxy at $z=9.25$ showing $\beta\sim-3$. Our major findings are summarized below:

\begin{itemize}
    \item By fitting $f(\lambda) \propto \lambda^\beta$ to the UV continua, we find EBG-1 showing $\beta=-2.99\pm0.15$, which is below the lower limit for no ionizing photon escape, $\beta=-2.6$, beyond the $2\sigma$ level. This small $\beta$ value does not change significantly by changing the fitting method and data reduction procedure, which suggest that the extremely blue UV continuum is not caused by systematics.

    \item The NIRSpec PRISM spectrum of EBG-1 shows [O \textsc{iii}] $\lambda\lambda$4959, 5007 emission lines. By estimating the SFR from the UV luminosity, we calculate the $L_\mathrm{[OIII]}/$SFR ratio for EBG-1. The comparison with the galaxies at $4<z<9$ compiled by \cite{Nakajima+2023} suggests that the [O \textsc{iii}] emission is slightly weaker than expected from the UV luminosity if $\fesc=0$ is assumed, indicating $\fesc\sim0.7$ for EBG-1.

    \item We compare the observed $\beta$ and EW(H$\beta$) with our Cloudy modeling. The extremely blue $\beta$ value imply $\fesc\gtrsim0.5$, which is consistent with the $\fesc$ value inferred from the [O \textsc{iii}] emission. However, the weak upper limit of H$\beta$ emission line prevents us from break the degeneracy between $\fesc$ and the shape of the ionizing source. It is thus important to conduct deeper spectroscopy on EBG-1.
\end{itemize}

\section*{Acknowledgments}
We are grateful to Steven L. Finkelstein, A. Ferrara, Moka Nishigaki, and Kuria Watanabe for the valuable discussions. This work is based on observations made with the NASA/ESA/CSA James Webb Space Telescope. The NIRSpec data of EBG-1 were obtained from the Mikulski Archive for Space Telescopes at the Space Telescope Science Institute, which is operated by the Association of Universities for Research in Astronomy, Inc., under NASA contract NAS 5-03127 for JWST. The data products presented herein were retrieved from DJA. DJA is an initiative of the Cosmic Dawn Center (DAWN), which is funded by the Danish National Research Foundation under grant DNRF140. We thank DJA for providing the reduced NIRSpec data. The observational data collected from DJA are associated with programs ERS 1345 (CEERS; PI: Finkelstein), DDT 2756 (PI: Chen), DDT 2750 (CEERS; PI: Arrabal Haro), GTO 1180 (PI: Eisenstein), GTO 1181 (PI: Eisenstein), GTO 1210 (PI: Luetzgendorf), GTO 1211 (PI: Isaak), GTO 1286 (PI: Luetzgendorf), DDT 6541 (PI: Egami), GO 1433 (PI: Coe),  GO 1747 (PI: Roberts-Borsani), GO 1810 (PI: Belli), GO 1871 (PI: Chisholm), GO 2110 (PI: Kriek), GO 2198 (PI: Barrufet), GO 2561 (UNCOVER; PI: Labbe), GO 2565 (PI: Glazebrook), DDT 2767 (PI: Kelly), ERO 2736 (PI: Pontoppidan),  GO 3215 (PI: Eisenstein), GO 4233 (PI: de Graaff) GO 4246 (PI: Abdurro'uf), DDT 4446(PI: Frye), and DDT 4557 (PI: Yan). The authors acknowledge the teams conducting these observations for publicly releasing the data. This publication is based on work supported by the World Premier International Research Center Initiative (WPI Initiative), MEXT, Japan, KAKENHI (20H00180, 21H04467, 21H04489, 24K07102) through the Japan Society for the Promotion of Science, and JST FOREST Program (JP-MJFR202Z). This work was supported by the joint research program of the Institute for Cosmic Ray Research (ICRR), University of Tokyo.

The JWST NIRSpec data presented in this article were obtained from the Mikulski Archive for Space Telescopes (MAST) at the Space Telescope Science Institute. The specific observations analyzed can be accessed via \dataset[doi: 10.17909/c19c-0a22]{https://doi.org/10.17909/c19c-0a22}.

\vspace{5mm}

\software{astropy \citep{astropy_2013, astropy_2018, astropy_2022}, NumPy \citep{Harris+2020_numpy}, matplotlib \citep{Hunter+2007_matplotlib}, SciPy \citep{Virtanen+2020_scipy}, corner \citep{Foreman-Mackey_2016_corner}
          Cloudy \citep{2013RMxAA..49..137F}, 
          emcee \citep{Foreman-Mackey+2013_emcee},
         msaexp \citep{msaexp}
          Prospector \citep{Johnson+2021_prospector}
          }

\bibliography{reference}{}
\bibliographystyle{aasjournal}

\end{document}